\documentstyle[11pt,moriond,epsfig]{article}

\def\be{\begin{equation}}
\def\ee{\end{equation}}
\def\bea{\begin{eqnarray}}
\def\eea{\end{eqnarray}}

\begin{document}
%\vspace*{1cm}
\title{Orthogonality Catastrophes in Carbon Nanotubes}

\author{Leon Balents }

\address{Bell Laboratories, Lucent Technologies,\\
Room 1D-368, 600 Mountain Ave., Murray Hill, NJ 07974}

\maketitle\abstracts{Carbon nanotubes provide a remarkably versatile
  system in which to explore the effects of Coulomb interactions in
  one dimension.  The most dramatic effects of strong
  electron-electron repulsion are {\sl orthogonality catastrophes}.
  These orthogonality catastrophes come in different varieties, and
  can be observed both in low-bias transport and tunneling
  measurements on nanotubes.  This article contains a review of
  previous work and new material on the crossover between Coulomb
  blockade and Luttinger behavior.}

\section{Introduction and Model}

In this section we describe the basic influence of interactions in
carbon nanotubes, following Ref.~1.  The band structure of
metallic nanotubes has been discussed by several
authors.\cite{Theory1,Theory2}\ It is well-captured by a simple
tight-binding model of $p_z$ electrons on the honeycomb lattice.  For
the metallic tubes, evaluating the resulting tight-binding band
structure for the discrete set of allowed quantized transverse momenta
$q_y$ leads to only two gapless {\sl one-dimensional} metallic bands
(for the simplest $(N,N)$ armchair tubes, these have $q_y =
0$).\cite{Theory1,Theory2}\ These dominate the low-energy physics,
disperse with the same velocity, $v_F$, and can be described by the
simple 1d free $ $Fermion model,
\begin{equation}
H_0 = \sum_{i,\alpha} \int dx v_F \left[
    \psi^{\dag}_{Ri\alpha}i\partial_x \psi^{\vphantom\dag}_{Ri\alpha}
    -\psi^{\dag}_{Li\alpha}i\partial_x \psi^{\vphantom\dag}_{Li\alpha}
  \right],  
\label{H_0}
\end{equation}
where $i=1,2$ labels the two bands, and $\alpha = \uparrow,\downarrow$
the electron spin.  We neglect curvature effects (which could open up
small gaps in all but the armchair tubes), since these are subdominant to
the Coulomb interactions.  Higher sub-bands (for the armchair tube
these have $q_y \neq 0$) are of course also present, and will be
returned to in Sec.~3.

We will make heavy use of the bosonized representation of
Eq.~\ref{H_0}, obtained by writing $\psi_{R/L;i\alpha} \sim
e^{i(\phi_{i\alpha} \pm \theta_{i\alpha})}$, where the dual fields
satisfy $[\phi_{i\alpha}(x), \theta_{j\beta}(y)] =
-i\pi \delta_{ij}\delta_{\alpha\beta} \Theta(x-y)$.  
Expressed in these variables (1) takes the form
$H_0 = \sum_{i,\alpha} {\cal H}_0(\theta_{i\alpha},\phi_{i\alpha})$
\begin{equation}
{\cal H}_0(\theta,\phi) = \int dx \, {v_F \over {2\pi}} [(\partial_x
\theta)^2 + (\partial_x \phi)^2].
\label{Hspin}
\end{equation}
The slowly varying electronic density in a given channel is given by
$\rho_{i\alpha} \equiv \psi_{Ri\alpha}^\dagger
\psi_{Ri\alpha}^{\vphantom\dag} + \psi_{Li\alpha}^\dagger
\psi_{Li\alpha}^{\vphantom\dagger} = \partial_x\theta_{i\alpha}/\pi$.
The normal modes of ${\cal H}_0$ describe long wavelength particle-hole 
excitations which propagate with a dispersion $\omega = v_F q$.

Turning to the interactions, a tremendous simplification occurs when
$N$ is large: the only couplings which survive in this limit are
{\it forward scattering} processes which involve small momentum
transfer.  Roughly speaking, this can be understood as follows.
``Interbranch" scattering processes (such as backscattering and
umklapp) involve a momentum transfer of order $2k_F \sim 1/a$, where
$a$ is the carbon-carbon bond length.  The matrix elements are
therefore dominated by the {\it short range} part of the interaction,
at distances $r\sim a$, where the interaction changes significantly
from site to site.  However, the electrons in the lowest sub-band are
spread out around the circumference of the tube, and for large $N$ the
probability of two electrons to be near each other is of order $1/N$.
For the Coulomb interaction, the resulting dimensionless interaction
vertices are of order $(e^2/h v_F) \times 1/N$\cite{Balents97,Twiston}.  By
contrast forward scattering processes, in which electrons stay in the
same branch, involve small momentum exchange.  They are dominated by the
{\it long range} part of the Coulomb interaction, at distances larger
than the radius, and there is no $1/N$ suppression.

For $N > 10$ it is thus appropriate to consider a {\it Luttinger
  model}, in which only forward scattering vertices are included.  A
further simplification arises because the {\it squared} moduli of the
electron wavefunctions in the two bands are {\it identical} and spin
independent.  All the forward-scattering vertices can thus be written
as a {\it single} interaction, coupling to the total charge density
$\rho_{\rm tot} = \sum_{i\alpha} \partial_x \theta_{i\alpha}/\pi$.

We will suppose that the Coulomb interaction is externally screened on 
a scale $R_s$, which is long compared to the tube radius $R$, but
short compared to the length of the tube.
For simplicity, we model this by a metallic cylinder of
radius $R_s$, placed around the nanotube.  From elementary
electrostatics, the energy to charge the
nanotube with an electron density $e\rho_{tot}$ is
\begin{equation}
  H_{\rm int} = e^2 \ln(R_s/R) \int dx  \rho_{\rm tot}^2  .
\end{equation}

Since $H_{\rm int}$ only involves $\rho_{\rm tot}$ it is
convenient to introduce a spin
and channel decomposition via, $\theta_{i,\rho/\sigma} =
(\theta_{i\uparrow} \pm \theta_{i\downarrow})/\sqrt{2} $ and
$\theta_{\mu \pm} = (\theta_{1\mu} \pm \theta_{2\mu})/\sqrt{2}$ with
$\mu = \rho, \sigma$, and similar definitions for $\phi$.  As defined,
the new fields $\theta_a$ and $\phi_a$ with $a=(\rho/\sigma, \pm)$,
satisfy the same canonical commutators $[\phi_a(x), \theta_b(y)] =
-i\pi\delta_{ab} \Theta(x-y)$.  
In the absence of interactions the Hamiltonian is simply
$H_0 = \sum_a \int_{x,\tau} {\cal H}_0(\theta_a,\phi_a)$. 
which describes three ``sectors" of neutral excitations and one
charged excitation.  Including the interactions only modifies
the charge sector, which is described by the sum of two terms
${\cal H}_\rho = {\cal H}_0(\theta_{\rho+},\phi_{\rho+}) 
+ H_{\rm int}(\theta_{\rho+})$ which may be written
\begin{equation}
{\cal H}_\rho = \int dx {v_\rho\over {2\pi }} \left[
  g^{-1}(\partial_x\theta_{\rho+})^2 
+ g  (\partial_x\phi_{\rho+})^2 \right].
\label{Hcharge}
\end{equation}
This describes the 1d accoustic {\it plasmon} which propagates with
the velocity \\
$v_\rho = \sqrt{v_F(v_F + (8e^2/\pi \hbar) \ln (R_s/R))}$ and is characterized
by the Luttinger parameter $g = v_F/v_\rho$.  Taking a screening
radius of $R_s = 0.1\mu$, one estimates $g \approx 0.2$.  

\section{Low-Bias Transport}

As evidenced in Eq.~\ref{Hcharge}, the elementary excitations of the
Luttinger liquid are collective modes, very different from the
quasiparticles of a normal Fermi liquid.  The most natural
experimental measure of this non-Fermi-liquid physics are therefore
those which probe the overlap between these excitations and bare
electrons.  The simplest way to achieve this is via a low-bias
transport experiment in which electrons are removed from a metallic source
lead and injected into the nanotube (and vice versa at the drain).
Surprisingly, for this purpose it 
is beneficial to have {\sl poor} contact between the leads and the
tube.  In this poorly-contacted limit, transport proceeds by
sequential tunneling of individual electrons into and out of the
nanotube, and is thus a measure of the single-particle overlap.  By
contrast, with ideal (adiabatic) contacts, the charge on the tube
is no longer a good quantum number, and transport occurs via a
collective flow of charge.  Despite the significant different between
the single-particle spectrum of Luttinger and Fermi liquids, their
collective mode spectra are quite similar.  Indeed, as discussed by
many authors, an ideally
contacted pure Luttinger liquid has a two-terminal resistance which
is completely unaffected by interactions.  The ideal
four-terminal resistance of a zero temperature Luttinger liquid is
{\sl zero}, a characteristic of ballistic transport.

We consider the transport through a single tunnel junction, which
could be between a normal lead and the nanotube, or between two
nanotubes.  The former situation has been investigated experimentally
and theoretically in Ref.~6 (see the paper by Bockrath for
details).  In general, the current through a single tunnel junction
can be determined perturbatively in the tunneling amplitude $w$.
Consider the tunneling Hamiltonian
\begin{equation}
  H_{\rm tunn.}(t) = \sum_\alpha w e^{i V t} c^\dagger_{l\alpha}
  c^{\vphantom\dagger}_{r\alpha}  +
  w^* e^{-iV t}c^\dagger_{r\alpha} c^{\vphantom\dagger}_{l\alpha}, 
\end{equation}
where $c_{r,l}^\dagger / c_{r,l}^{\vphantom\dagger}$
create/annihilate electrons to the left and right of the junction.  If 
the right-hand-side is a nanotube, there is in general some tunneling
amplitude into each of the Dirac channels: $c_{r\alpha} =
\sum_{i=1,2}\sum_{P=R/L} \Phi^{(r)}_{P i} \psi_{Pi\alpha}$.  We have
chosen to include the voltage difference between the two leads here as 
a time-dependent vector potential.  The current across the junction
follows from gauge invariance: $I = \partial {\cal L}/\partial a_x$,
where $a_x$ is the component of the vector potential normal to the
junction.  The first non-zero contribution to the current in
time-dependent perturbation theory in $w$ is 
\begin{equation}
  \langle I(t) \rangle \propto |w|^2 \int_{-\infty}^t e^{iV(t-t')} {\cal 
    A}_{\rm ret.}(t-t') \equiv {\cal A}_{\rm ret.}(V),
  \label{tunnel}
\end{equation}
where ${\cal A}_{\rm ret.}$ is the standard retarded correlation
function of the operator $\hat{A} = c_{l\alpha}^\dagger
c_{r\alpha}^{\vphantom\dagger}$.  Since the overall magnitude of the
current depends on an unknown transmission probability, we will
simplify various formulae by omitting constant prefactors where
possible. The retarded correlator is obtained in the usual way by
analytic continuation from the imaginary-time correlator ${\cal
  A}(\tau) = \langle T_\tau A^{\vphantom\dagger}(\tau)
A^\dagger(0)\rangle$.  In particular,
\begin{equation}
  {\cal A}_{\rm ret.}(V) = \sinh (\beta V/2) \int_{-\infty}^\infty \!
  dt\, e^{-iVt} {\cal A}({\beta \over 2} - it),
  \label{analytic}
\end{equation}
where, as usual $\beta = 1/k_B T$.

\subsection{Thermodynamic Limit}

We first consider the case in which the systems on either side of the
contact are semi-infinite.  This is a valid assumption for a finite
Luttinger liquid provided the temperature or bias voltage is large
compared to its capacitive charging energy.  For both the case of
Fermi-liquid to nanotube and nanotube--nanotube junctions, the
imaginary-time correlator above takes the form
\begin{equation}
  {\cal A}(\tau) = \left( {{\pi \over \beta} \over \sin{{\pi \tau
          \over\beta}}} \right)^{2+\alpha},
  \label{acft}
\end{equation}
with $\alpha>0$, which can be obtained straightforwardly from the
bosonization formulae in Sec.~1.  This form obtains in many distinct
cases -- e.g. bulk contact between a Fermi liquid and a nanotube, bulk
contact between two nanotubes, and end-to-end contact between two
nanotubes.  The different cases are distinguished by the value of the
exponent $\alpha$.  To get a feeling for the physical meaning of
$\alpha$, imagine a Fermi's golden rule estimate of the current across
the junction.  If the initial state is the ground state, the final
state consists of e.g. a hole in the left lead and an electron in the
right.  Fermi's golden rule gives a density of states factor which is
the product of the electron density of states on the right and the
hole density of states on the left.  Indeed, a direct calculation of
the tunneling density of states $\rho^{\rm tun.}_{r/l}$ for a Luttinger
liquid gives the formula $\alpha = \alpha_l + \alpha_r$, where
$\rho^{\rm tun.}_{r/l}(\epsilon) \sim |\epsilon|^{\alpha_{r/l}}$.  

The density of states exponent depends crucially on the Luttinger
parameter $g$, and also on the point of contact.  In particular, 
\begin{equation}
  \alpha =
  \cases{(g^{-1}-1)/4 & near cap \cr (g+g^{-1}-2)/8 & in bulk }
\end{equation}
Note that, since $g<1$, the orthogonality exponent $\alpha$ is
significantly larger near the end of a nanotube.  This enhanced
interaction effect arises essentially from the decreased ability of an
added charge to spread away from the cap area relative to the bulk.
A contact may be considered an end contact if the distance from the end
of the nanotube is smaller than the lesser of the characteristic
length scales $L_T = \hbar v_F/k_B T, L_V = \hbar v_F/eV$.  It is
important to realize that this is not a small effect: for $g \ll 1$,
the orthogonality exponent near the cap is nearly {\sl twice} that in
the bulk. 

Using Eq.~\ref{acft}\ and Eq.~\ref{analytic}, one obtains the $I$--$V$ 
curve
\begin{equation}
  I = I_0 T^{1+\alpha} \sinh (\beta V/2) \left| \Gamma\left( 1 +
      {\alpha \over 2} + i {\beta V \over {2\pi}}\right)\right|^2,
  \label{IV}
\end{equation}
where $I_0 \propto |w|^2$ is an unknown prefactor and $\Gamma(z)$ is
the Gamma function.  This form is well-known from the literature on
tunneling and Caldeira--Leggett models.  A slightly more complicated
formula can be obtained for the differential conductance $G = dI/dV$
by differentiation;  differentiating only the $\sinh (\beta V/2)$
prefactor gives a common approximate result (as the remainder is small 
at low biases).  Eq.~\ref{IV}\ has an important {\sl scaling}
property.  The quantities $I/T^{1+\alpha}$ and $G/T^\alpha$ are
functions only of the ratio $V/T$.  This implies that IV curves taken at
different temperatures should collapse if, e.g. $G/T^\alpha$ is
plotted as a function of $V/T$.  See the paper by Marc Bockrath in
this volume for an experimental observation of this behavior.

\subsection{Charging Effects}

Coulomb blockade is a ubiquitous phenomena in mesoscopics physics.
For a few micron-long nanotube, the capacitive charging energy can be
of the order of tens of Kelvin, and affects the transport dramatically 
at low temperatures.  The crossover between Luttinger liquid and
Coulomb blockade behavior is a complex and relatively unstudied
problem.  We will discuss a simple model of charging effects for a
single tunnel junction between two nanotubes.  The charging behavior
is, unfortunately, much less universal that the Luttinger liquid
results of the previous subsection, so we must make a number of
assumptions to progress.

A panoply of energy scales exist in a finite Luttinger liquid.  The
largest is generally the charging energy.  In the Luttinger model,
this is the energy of a {\sl zero mode} in which each of the phase
fields $\theta_a$ winds by $\pm \pi/2$.  For $g\ll 1$, this is
dominated by the charge mode $\theta_\rho$, and from
Eqs.~\ref{Hspin},\ref{Hcharge}, $E_C \approx \pi \hbar
v_F/8g^2 L \approx (e^2/L)\ln R_s/R$.  For a one-dimensional system,
the level spacing also scales inversely with the length.  In
particular, the Luttinger liquid exhibits {\sl two} such level
spacings due to spin-charge separation: a plasmon energy
$\varepsilon_\rho = \pi \hbar v_\rho/L$ and a single-particle energy
$\varepsilon_0 = \pi\hbar v_F/L$.  For $g\ll 1$, one has $E_C \gg
\varepsilon_\rho \gg \varepsilon_0$.  We will therefore consider only
the largest of these energies, $E_C$.  This corresponds physically to
including charging effects but not individual level quantization.

Within this charging-only model, we will consider the specific case of
a single tunnel-junction between two finite-length nanotubes.  The
treatment here is valid provided that the resistance of the
nanotube-nanotube junction is much larger than the resistance to the
leads.  In this case, at least for not too low temperatures, transport
is dominated by the internal tunnel junction.  We require $R_{\rm
  junction} \gg R_{\rm leads} \gg h/e^2$.

To proceed, we must specify the capacitances in the system.  In
general, the charging energy is given by an energy function
$E_{n_1,n_2}$, where $n_1$ and $n_2$ are the charges (in units of $e$)
of the left and right sides of the system.  Given such a function,
Eq.~\ref{analytic}\ can still be applied, but ${\cal A}(\tau)$ must be
recalculated.  Because the charges are zero-mode quantities, the
statistical average simply factors into a product of a zero-mode
contribution and the thermodynamic contribution in Eq.~\ref{acft}.
Inserting into Eqs.~\ref{tunnel}-\ref{analytic}, one obtains
\begin{equation}
  I = I_0 {T^{1+\alpha} \over Z} \sinh \left( \beta V/2\right)
  \sum_{n_1, n_2} e^{-\beta E^+_{n_1,n_2}} \left| \Gamma\left( 1 +
      {\alpha \over 2} + i {\beta(V-\Delta E_{n_1,n_2}) \over
        {2\pi}}\right)\right|^2,  
  \label{master}
\end{equation}
where the ``partition function'' is $Z = \sum_{n_1,n_2} e^{-\beta
  E_{n_1,n_2} }$. We have also defined the energies $E^+_{n_1,n_2}  =
{1 \over 2}\left( E_{n_1+1,n_2-1} + E_{n_1,n_2}\right)$, 
  $\Delta E_{n_1,n_2}  =  E_{n_1+1,n_2-1} - E_{n_1,n_2}$.
In general, a quadratic energy function $E_{n_1,n_2}$ contains five
non-trivial parameters.  It is helpful to consider a simplified form,
\begin{equation}
  E_{n_1,n_2} = U(n_1+n_2 -\overline{n})^2 +
  W(n_1-n_2-\overline{m})^2.
  \label{simple}
\end{equation}
Here $W$ can be viewed as a capacitive charging energy for the
junction itself, while $U$ gives an ``bulk'' contribution.  A gate
potential is included via $\overline{n},\overline{m}$.  Changing a
single gate potential corresponds to motion along some straight line
in the $\overline{n},\overline{m}$ plane.
\begin{figure}
\hspace{1.8in}\epsfxsize=3.0in\epsffile{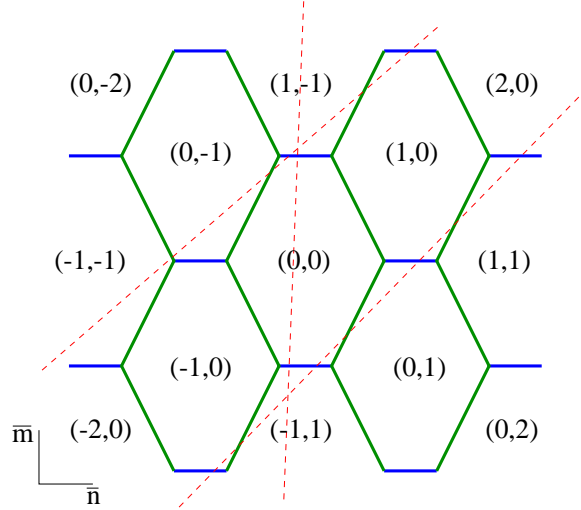}
\caption{Ground state charge configurations in the
  $\overline{n}$--$\overline{m}$ plane, assuming $W<U$.  Inside each
  hexagon the charge state is indicated by the ordered pair
  $(n_1,n_2)$.  For $W>U$, the pattern
  is rotated $90$ degrees and there are {\sl no} horizontal segments.
  Some possible trajectories as the gate voltage is varied
  are indicated schematically by the dashed red lines.}
\end{figure}
Eqs.~\ref{master}-\ref{simple}\ encompass a tremendously rich
behavior.  Consider the Coulomb blockade regime at low temperature.
At zero temperature, the ground state charge configurations of the
system (for $w=0$) are shown in Fig.~1.  As
$\overline{m},\overline{n}$ are varied, the system makes two types of
transitions between different charge states.  Across the diagonal
lines, one electron is added or removed from either the left or right
nanotube, changing the net charge (and spin) on both tubes.  Across the
horizontal lines, one electron is transferred from the right to the
left tube or vice versa, leaving the total change unchanged.  In the
limit we are considering, transport is dominated by tunneling across
the inter-tube junction.  Peaks in the zero-bias conductance thus
occur only across the horizontal lines.  

As an external gate potential is varied, the system traces out a
straight line in the $\overline{m}$--$\overline{n}$ plane.  The
particular slope and intercept of this line is determined by the
precise geometry of the tubes and the gate.  The system will thus
generically undergo a sequence of ground-state transitions, some of
which (crossing the diagonal lines) do not give rise to conductance
peaks, but which change the charge and spin.  Such ``internal''
transitions have indeed been observed in nanotubes.\cite{Tans98}\ 
Moreover, depending upon the particular path taken, the locations of
Coulomb blockade peaks as a function of gate voltage (crossings of
horizontal lines) can be quite complex, e.g. quasiperiodic or
containing a fairly regular series of peaks followed by a region of
zero conductance.

Let us return to the problem of the crossover between Coulomb blockade
and Luttinger behavior, contained in Eqs.~\ref{master}--\ref{simple}.
A general analysis is beyond the scope of these proceedings.  We
therefore content ourselves with a few simple situations.  The most
``universal'' limit is the high-temperature case, basically when $T >
W$ (note that $W$ and not $U$ sets the size of the energy that an
electron must obtain to cross the tunnel junction).  In this case, the
sum in Eq.~\ref{master}\ can be approximated by an integral, and
\begin{equation}
  I \approx I_0 2^{-\alpha} \pi \Gamma(2+\alpha) T^{1+\alpha} \sinh(
  \beta V/2) e^{-\beta W} \int_0^\infty \! dt\, {{e^{-4\beta W t^2}
      \cos \beta V  t} \over {(\cosh \pi t)^{2+\alpha}}}.
  \label{highT}
\end{equation}
At very high temperatures, the exponential terms in $W$ are negligible 
and the result reduces to Eq.~\ref{IV}.  An appealing feature of
Eq.~\ref{highT}\ is its independence of $\overline{m},\overline{n}$
and $U$.  

Unfortunately, the low-temperature limit is much less universal.  As
an illustration of the application of Eqs.~\ref{master}-\ref{simple},
consider the behavior of the zero-bias conductance $G(T)$ as a
function of temperature when the gate voltage is tuned to a zero-bias
conductance peak.  In this case, one finds power-law behavior at {\sl
  both} low and high temperatures: $G(T) \sim G_0(W/T) T^\alpha$.  The 
prefactor, however, has the amusing property that $G_0(W/T \gg 1) =
G_0(W/T \ll 1)/2$.  The factor of two comes physically
from the fluctuations of the system between two generate ground states
at low temperatures, and reflects the zero-temperature
entropy of the two degenerate ground states of the system!  Amusingly, 
a more detailed examination shows that $G(T)/T^\alpha$ actually
crosses non-monotonically between these limits.\cite{unpub}\

\section{High-Bias Tunneling Spectra}

We have seen in the previous section how Coulomb interactions give
rise to numerous interesting non-linearities in low-bias tunneling
transport.  Tunneling spectra also provide a powerful probe of
high-energy excitations.  In nanotubes, tunneling experiments have
verified the presence of van Hove singularities in the density of
states (at $eV$ energies) due to the higher
sub-bands.\cite{Wildoer98,Odom98}\  In non-interacting band theory, these
singularities are of inverse square-root form, giving a contribution
$\rho_0(\epsilon) \sim \sqrt{m} (\epsilon-\Delta)^{-1/2}
\Theta(\epsilon-\Delta)$ for energies just above the subband edge at
$\epsilon=\Delta$ ($m$ is the subband effective mass).

How do interactions affect these van Hove singularities?  A
simplified, though unphysical model in which the mass of the higher
subband is taken to be infinite provides considerable insight.  In
this limit the higher energy subbands can be replaced by discrete,
localized levels.  The ``x-ray edge'' problem of a localized level
interacting with a conduction sea was solved by Nozieres and de
Dominicis\cite{Nozieres69}, and is one of the first demonstrations of
an orthogonality catastrophe.  Physically, the core hole is
``dressed'' through interactions with conduction electrons, which see
the hole as a scattering center.  This leads to a broadening and
reduction of the tunneling density of states from a sharp
delta-function to a power law singularity.

Remarkably, these x-ray edge effects persist even for this finite mass
case, as we now proceed to demonstrate.\cite{Ogawa92,heavy}\ We argue that
a necessary and sufficient condition for the presence of such
finite-energy singularities is a conserved quantum number
distinguishing the states of the higher subband from the conduction
states.  In the case of the carbon nanotube, this is an angular
momentum quanta.  If such a distinguishing quantum number is absent
(as might occur in a nanotube due to breaking of the rotational
symmetry by interactions with a substrate), we expect the van Hove
peak to be rounded and rendered completely nonsingular.

We describe here a simple forward-scattering model of the interaction
of the conduction electrons with the higher subband -- other
interaction channels are discussed in Ref.~13.  Near the
putative van Hove singularity, the unoccupied 1d subband can be described by a
non-relativistic electron operator $d,d^\dagger$:
\begin{equation}
  H_0^d = \int \! dx\, d_\alpha^\dagger \left[ -{1 \over
      {2m}}\partial_x^2 + \Delta\right] d^{\vphantom\dagger}_\alpha.
  \label{eq:heavy}
\end{equation} 
Here $\Delta$ is the gap to the first subband and $m$ is an effective
mass.  The electron field satisfies
$\{d_\alpha^{\vphantom\dagger}(x),d_\beta^\dagger(x')\} =
\delta_{\alpha\beta}\delta(x-x')$.  In the case of a carbon nanotube,
there are actually multiple degenerate subbands at energy $\Delta$.
This degeneracy is unimportant within the forward-scattering model, as
the tunneling DOS involves only states with a single excited electron.

The interaction Hamiltonian is $H_{\rm int} =
\int_x {\cal H}_{\rm int}$, with
\begin{equation}
  {\cal H}_{\rm int} = e^2 \ln(R_s/R) \left[ \left( d^\dagger
      d^{\vphantom\dagger} \right)^2 + {4 \over \pi}
    \partial_x\theta_{\rho+} d^\dagger d^{\vphantom\dagger} \right].
  \label{hint}
\end{equation}

To understand the effects of ${\cal H}_{\rm int.}$, 
%it is helpful to
%absorb the Hamiltonian density into Eq.~\ref{Hcharge}, viz.
%\begin{equation}
%  {\cal H} = {v_\rho \over {2\pi g}}\left[\partial_x\theta_\rho +
%    \gamma d^\dagger d^{\vphantom\dagger} 
%  \right]^2 + {{gv_\rho} \over {2\pi}}(\partial_x\phi)^2 + {\cal
%    H}_0^d,
%  \label{eq:bosonmodel}
%\end{equation}
%where the coupling constant $\gamma = {\pi \over 2}(1-g^2)$.
consider the canonical transformation
\begin{eqnarray}
  \theta_{\rho+}(x) & = & \tilde\theta_{\rho+}(x) - \gamma
  \int^x_{-\infty} \! dx' 
  \, d^\dagger\!(x') d^{\vphantom\dagger}(x'), \label{eq:densityshift}\\
  d(x) & = & e^{i\gamma\phi_{\rho+}(x)/\pi} \tilde{d}(x), \label{eq:string}
\end{eqnarray}
where $\gamma = \pi(1-g^2)/2$.
Eqs.~\ref{eq:densityshift}-\ref{eq:string}\ embody the physical
process in which the conduction sea {\sl adiabatically adjusts to the
  heavy particle}.   In particular, Eq.~\ref{eq:densityshift}\
represents the depletion of the conduction electron density near the
heavy particle due to Coulomb repulsion.  Eq.~\ref{eq:string}\
represents phase shifts of these conduction electrons when the heavy
particle is introduced.  Formally, the exponential of the dual
($\phi$) field in Eq.~\ref{eq:string}\ is a Jordan-Wigner ``string''
operator which has been attached to the heavy particle.

Remarkably, although the canonical transformation,
Eqs.~\ref{eq:densityshift}-\ref{eq:string}, does not remove the
interaction completely, it does transform the Hamiltonian into one
with only {\sl irrelevant} couplings in the renormalization group
sense.\cite{heavy}\ This indicates that at long times and distances,
the transformed fermion and boson correlation functions asymptotically
factorize.  From this result, one straightforwardly obtains a modified
van Hove singularity, gives a modified van Hove singularity:
\begin{equation}
  \rho(\epsilon) \sim \rho_0 \left( \Delta \over {\epsilon-\Delta}
  \right)^{{1 \over 2} - \beta} \Theta(\epsilon-\Delta),
  \label{dos}
\end{equation}
where $\Theta(x)$ is the heavyside step function, and the
orthogonality exponent $\beta = (1-g^2)^2/(8 g) \approx 0.3$ for
typical metallic nanotubes.

\section*{Acknowledgments}
I would like to thank Charlie Kane and Matthew Fisher for enjoyable
and insightful collaboration on the early stages of this work.

\end{document}